\documentclass[11pt]{article}
\usepackage{graphicx}
\usepackage{amssymb}
\usepackage{amsmath}
\usepackage{booktabs}
\usepackage{subcaption}
\usepackage{tikz}
\usepackage{url}
\renewcommand{\title}[1]{{\noindent\large\bfseries#1\medskip\\}}
\renewcommand{\author}[2]{{\noindent #1 \medskip\\ \small #2 \medskip\\}}
\usepackage[letterpaper,margin=20mm]{geometry}
\usepackage{csquotes}
\usepackage{svg}
\usepackage{xpatch} 

\usepackage[colorlinks=true, linkcolor=blue, citecolor=blue, urlcolor=blue]{hyperref}

\usepackage[
  backend=biber,
  bibstyle=apa,          
  citestyle=numeric-comp,
  sorting=none
]{biblatex}

\DeclareLanguageMapping{american}{american-apa}
\DeclareFieldFormat{labelnumberwidth}{\mkbibbrackets{#1}}
\defbibenvironment{bibliography}
  {\list
     {\printfield[labelnumberwidth]{labelnumber}}
     {\setlength{\labelwidth}{\labelnumberwidth}%
      \setlength{\leftmargin}{\labelwidth}%
      \setlength{\labelsep}{\biblabelsep}%
      \addtolength{\leftmargin}{\labelsep}%
      \setlength{\itemsep}{\bibitemsep}%
      \setlength{\parsep}{\bibparsep}}}
  {\endlist}
  {\item}

\begin{document}

\title{Evaluation of Time Series Forecasting Models for Predicting Lung
Cancer Mortality Rates in the United States: A Comparison with Altuhaifa
(2023) Study.}

\author{Emmanuel Kubuafor\textsuperscript{1*}, Dennis Baidoo\textsuperscript{1}, Onyedikachi Joshua Okeke\textsuperscript{1}, Robert Amevor\textsuperscript{1}, Godfred Arhin\textsuperscript{2} and Joshua Tetteh Korley\textsuperscript{3}
}
{
\textsuperscript{1} Department of Mathematics and Statistics, University of New Mexico, Albuquerque, NM, USA. \\ 
\textsuperscript{2} Department of Mathematical Sciences, The University of Texas at El Paso, El Paso, TX, USA. \\ 
\textsuperscript{3} Department of Statistics, Western Michigan University, Kalamazoo, MI, USA. \\

*Corresponding author: \texttt{ekubuafor28@gmail.com}
}

\noindent \textbf{{\Large Abstract}}\\

\noindent This paper evaluates the performance of the following time series forecasting models - Simple Exponential Smoothing (SES), Holt's Double Exponential Smoothing (HDES), and Autoregressive Integrated Moving Average (ARIMA) - in predicting lung cancer mortality rates in the United States. It builds upon the work of Altuhaifa, which used Surveillance, Epidemiology, and End Results (SEER) data from 1975-2018 to evaluate these models. Altuhaifa's study found that ARIMA (0,2,2), SES with smoothing parameter $\alpha=0.995$, and HDES with parameters $\alpha=0.4$ and $\beta=0.9$ were the optimal models from their analysis, with HDES providing the lowest Root Mean Squared Error (RMSE) of 132.91. The paper extends the dataset to 2021 and re-evaluates the models. Using the same SEER data from 1975-2021, it identifies ARIMA (0,2,2), SES ($\alpha=0.999$), and HDES ($\alpha=0.5221$, $\beta=0.5219$) as the best-fitting models. Interestingly, ARIMA (0,2,2) and HDES yield the lowest RMSE of 2.56. To obtain forecasts with higher accuracy, an average model (HDES-ARIMA) consisting of HDES and ARIMA was constructed to leverage their strengths. The HDES-ARIMA model also achieves an RMSE of 2.56. The forecast from the average model suggests declining lung cancer mortality rates in the United States. The study highlights how expanding datasets and re-evaluating models can provide updated insights. It recommends further analysis using monthly data separated by gender, ethnicity, and state to understand lung cancer mortality dynamics in the United States. Overall, advanced time series methods like HDES and ARIMA show strong potential for accurately forecasting this major public health issue.


\section{\Large Introduction}

\noindent The third most common cancer in the United States following skin cancer and breast cancer is lung cancer \cite{cdcLungCancer}, hence the need for strategic interventions against this fatal disease. Despite being the third most common form of cancer, it accounted for the most (approximately 1.8 million) cancer-related deaths in 2020 \cite{who2024}. By leveraging SEER data on lung cancer mortality rates, models can be developed to forecast lung cancer mortality for effective public health decision-making. Accurate forecasts assist healthcare providers in addressing patient requirements, optimizing hospital resources, and planning intervention measures effectively \cite{bhati2023}. A previous study by Altuhaifa (2023) demonstrated the effectiveness of ARIMA, SES, and HDES models in forecasting lung cancer mortality. However, the performance of these models may vary depending on the dataset length, mortality trend shifts, and computational methodologies. This study builds upon Altuhaifa's study \cite{altuhaifa2023} by extending the dataset through 2021 and conducting a comparative evaluation of SES, HDES, and ARIMA models. By doing so, it seeks to identify the most suitable model for forecasting lung cancer mortality in the United States while addressing potential limitations in previous analyses. Additionally, this study explores the advantages of combining HDES and ARIMA models into an integrated approach to enhance forecast accuracy. By filling this research gap, the study aims to provide more precise mortality projections that can guide public health interventions and policy decisions.

\section{\Large Methodology}

\noindent Altuhaifa (2023) used SEER 8 registry data on mortality rates of lung cancer including 545,486 patients \cite{altuhaifa2023}. The SEER 8 data covers eight U.S. regions - Connecticut, Detroit, Hawaii, Iowa, New Mexico, San Francisco-Oakland, Seattle-Puget Sound, and Utah - providing diverse insights into cancer trends across urban, rural, and demographically unique populations. This regional diversity allows SEER 8 to capture valuable insights into cancer incidence, survival, and mortality across various racial, ethnic, and socioeconomic groups, supporting comprehensive national cancer research. Employing Python 3.9 with Anaconda 2022.10, ARIMA, SES, and HDES models were constructed. For the ARIMA model, parameters (p,d,q) were selected by using autocorrelation and partial autocorrelation plots to determine potential values for p and q, while the differencing parameter d was identified by iteratively differencing the data until stationarity was achieved, as confirmed by the augmented Dickey-Fuller test. The Akaike Information Criterion (AIC) was then used to evaluate and select the optimal ARIMA model configuration. The SES model's smoothing parameter $\alpha$, and the HDES model's parameters $\alpha$ and $\beta$ were optimized via grid search. The evaluation of the models involved partitioning the data into 75\% and 25\%, train and test sets respectively; and assessing the performance of these models using Mean Absolute Percentage Error (MAPE), Mean Absolute Error (MAE), and RMSE metrics.

\noindent This paper utilized U.S. SEER 8 data on lung cancer mortality rates from 1975 to 2021 \cite{cancerCancerLung}. The data was split into training (1975-2012) and test (2013-2021) sets. Using R software version 3.4.2, several ARIMA models were constructed on the training data and the best model was selected based on AIC and residual assumptions of white noise, normality, and autocorrelation satisfaction. Generally, the ARIMA model can mathematically be represented as:

\begin{equation}
y_{t}^{'} = c + \phi_{1}y_{t - 1}^{'} + \cdots + \phi_{p}y_{t - p}^{'} + \theta_{1}\epsilon_{t - 1} + \cdots + \theta_{q}\epsilon_{t - q} + \epsilon_{t}
\end{equation}

\noindent where $y_{t}^{'}$ is the differenced time series at time $t$, $c$ is a constant term which represents the mean level of the stationary process, $\phi_{1},\phi_{2},\ldots,\phi_{p}$ are the autoregressive coefficients for the lagged differenced values for the time series $y_{t - 1}^{'},\ y_{t - 2}^{'},\ldots,\ y_{t - p}^{'}$, and $\theta_{1},\theta_{2},\ldots,\theta_{q}$ are the coefficients for the lagged error terms $\epsilon_{1},\epsilon_{2},\ldots,\epsilon_{q}$. Our approach to the ARIMA modeling encompassed a two-stage process; visualization and parametric optimization. The initial phase focused on visualization, where we constructed line plots to understand the underlying patterns in lung cancer mortality rates. This visual analytics approach proved crucial for preliminary pattern recognition.

\noindent One fundamental challenge in our analysis was addressing the issue of non-stationarity. Traditional ARIMA time series analysis demands that the series exhibit mean and variance stability across its domain. Our data initially displayed non-stationarity, a common characteristic in medical time series data, and we needed to transform it to make it stationary.

\noindent To confirm this instability, we implemented the augmented Dickey-Fuller (ADF) test, which provides a statistical framework for assessing the stationarity of the series. The ADF test hypothesis framework is structured as: ($H_0$: Series is not stationary vs $H_1$: series is stationary). The statistical significance for this test was evaluated at $\alpha = 0.05$ where p-values less than this threshold confirmed non-stationarity.

\noindent To address this issue of non-stationarity, we used a progressive differencing framework, implementing two levels of transformation. This method involves subtracting prior time series values from the present observations. In first-order differencing, each observation is subtracted from the most recent previous value in the series, as shown in \textbf{Equation 2}. Second-order differencing applies this same process twice, as seen \textbf{in Equation 3}.

\begin{equation}
y_{t}^{'} = y_{t} - y_{t - 1}
\end{equation}

\begin{equation}
y_{t}^{''} = y_{t} - 2y_{t - 1} + y_{t - 2}
\end{equation}

\noindent $y_{t}^{'}$ and $y_{t}^{''}$ represents differenced series given first-order and second-order differencing respectively. $y_{t}$ denotes the mortality rate at time t with subscripts $t - 1$ and $t - 2$ indicating lagged observations. After verifying stationarity through an Augmented Dickey-Fuller (ADF) test, we then proceeded to estimate the appropriate ARIMA (Autoregressive Integrated Moving Average) model parameters. The ARIMA (p, d, q) model consists of three key components: the lag order (p), the degree of differencing (d), and the order of moving average (q). To determine the optimal ARIMA model parameters, we used the Akaike Information Criterion (AIC), which assesses model fit using maximum likelihood estimates.

\noindent The SES model, suitable for forecasting data with no clear trend or seasonal pattern was the second model considered. The forecast equation for the Simple Exponential Smoothing model can mathematically be represented as:

\begin{equation}
\hat{y}_{t + h|t} = l_{t}
\end{equation}

\begin{equation}
l_{t} = \alpha y_{t} + (1 - \alpha)l_{t - 1}
\end{equation}

\noindent where $\hat{y}_{t + h|t} = l_{t}$ is the forecast at time $t + h$ given $t$, $l_{t}$ is the level of the series at time $t$, $\alpha$ is the smoothing parameter, and $y_{t}$ is the observed value at time $t$.

\noindent The accuracy of SES forecasts relies heavily on the selection of the smoothing parameter, denoted as $\alpha$. The smoothing parameter $\alpha$ influences how much weight is given to the most recent observation versus the previous forecast. Values of $\alpha$ closer to 1 place more emphasis on the latest data point, while values nearer to 0 give more weight to the previous forecast. The estimated value of $\alpha$ that makes the model optimal is derived based on a grid search.

\noindent Holt's Double Exponential Smoothing (HDES) technique is a time series forecasting method designed to capture trends in data without accounting for seasonality. HDES is an extension of Simple Exponential Smoothing, which applies two smoothing parameters $\alpha$ and $\beta$ for the data components: level and trend respectively. The HDES model can mathematically be represented as:

\begin{equation}
\hat{y}_{t + h|t} = l_{t} + hb_{t}
\end{equation}

\begin{equation}
l_{t} = \alpha y_{t} + (1 - \alpha)\left( l_{t - 1} + b_{t - 1} \right)
\end{equation}

\begin{equation}
b_{t} = \beta^{*}\left( l_{t} - l_{t - 1} \right) + \left( 1 - \beta^{*} \right)b_{t - 1}
\end{equation}

\noindent where $b_{t}$ denotes an estimate of the trend (slope) at time t, and $\beta^{*}$ is the smoothing parameter of the trend, where $(0 < \beta^{*} < 1)$.

\noindent The parameters $\alpha$ and $\beta$ are estimated by minimizing the sum of squared errors between observed values and forecasts over the historical data using the method of maximum likelihood estimation (MLE), which automatically adjusts $\alpha$ and $\beta$ to minimize forecast errors. HDES is particularly useful for data with a clear trend but no seasonal patterns and is computationally efficient, making it well-suited for short-term to medium-term forecasting. Additionally, the residuals of the SES and HDES models were also examined for model appropriateness.

\noindent Based on the research of Altuhaifa (2023), the HDES model was the best model for forecasting lung cancer mortality \cite{altuhaifa2023}. A previous study by Alzhrani (2016) showed that the ARIMA model was the best for forecasting lung cancer mortality \cite{alzahrani2016}. Furthermore, Bates \& Clive (1969) demonstrated that combining forecasts frequently improves forecast accuracy \cite{bates1969}. Twenty years later, Clemen (1989) also wrote that forecast accuracy increases when many forecasts are combined \cite{clemen1989}. This infers that by merely averaging the forecasts from suitable models, one can frequently get significant forecast accuracy improvements. To leverage the forecasting strength of the ARIMA and HDES models, an average model (HDES-ARIMA) was constructed. Cross-validation utilizing the test data was performed and the models were compared using evaluation metrics consisting of RMSE, MAPE, and MAE.

\section{\Large Results}

\subsection{Altuhaifa (2023)}

\noindent The analysis made by Altuhaifa (2023) generated the optimal ARIMA model, ARIMA (0, 2, 2) having the lowest AIC value and satisfying the residual assumptions of white noise and normality \cite{altuhaifa2023}. The SES model smoothing parameter $\alpha$ optimized via grid search produced a value of 0.995 with the HDES model's parameters $\alpha$ and $\beta$ having values of 0.4 and 0.9 respectively as optimized values. The forecasting performance of each of the three models was assessed using the test data. The HDES model was selected as the best forecast model for lung cancer mortality, producing the lowest RMSE of 132.91. The results from the studies are summarized in Table \ref{Tab:1} below.

\begin{table}[h!]
\centering
\caption{Evaluation Metrics for SES, HDES, and ARIMA Models Using Test Data (2008-2018).}
\label{Tab:1}
\begin{tabular}{lccc}
\toprule
Models & MAPE & MAE & RMSE \\
\midrule
ARIMA (0, 2, 2) & 0.0120 & 117.700 & 144.360 \\
SES ($\alpha$ = 0.995) & 0.0428 & 380.454 & 541.187 \\
HDES & 0.0114 & 106.700 & 132.910 \\
\bottomrule
\end{tabular}
\end{table}

\subsection{Present Study}

\noindent This study extends the period of Altuhaifa's study by using data from 1975 to 2021. Figure \ref{fig:1} below, which is a plot of lung cancer mortality per 100,000 against years, depicts that lung cancer mortality rates in the USA gradually increased from 1975 to 1996 and then started to decrease.

\begin{figure}[htbp!]
\centering
\includegraphics[width=0.5\textwidth]{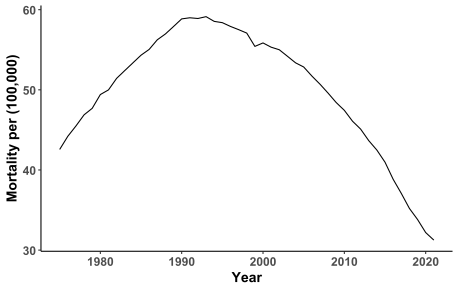}
\caption{Lung Cancer Mortality Rate in the USA}
\label{fig:1}
\end{figure}

\begin{figure}[htbp!]
\centering
\includegraphics[width=0.5\textwidth]{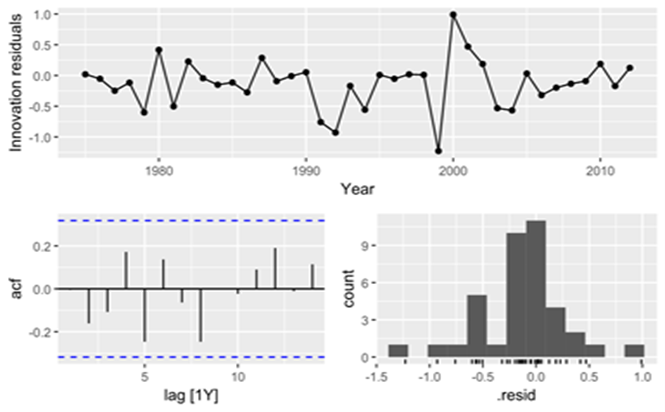}
\caption{Residual plot of ARIMA (0,2,2)}
\label{fig:2}
\end{figure}

\noindent Among the ARIMA models tested, ARIMA (0,2,2) was optimal, with the lowest AIC and residuals meeting model assumptions. The ACF plot shows no significant autocorrelation, while residual plots indicate normality, as seen in Figure \ref{fig:2}. Statistical tests confirm the absence of deviations: Box-Ljung (white noise, p-value = 0.4693), Shapiro-Wilk (normality, p-value = 0.1206), and KPSS (stationarity, p-value = 0.1).

\begin{figure}[htbp!]
\centering
\includegraphics[width=0.5\textwidth]{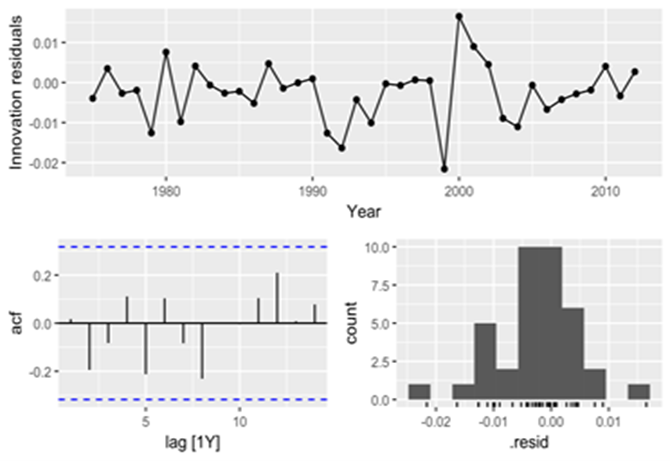}
\caption{Residual plot of HDES}
\label{fig:3}
\end{figure}

\begin{figure}[htbp!]
\centering
\includegraphics[width=0.5\textwidth]{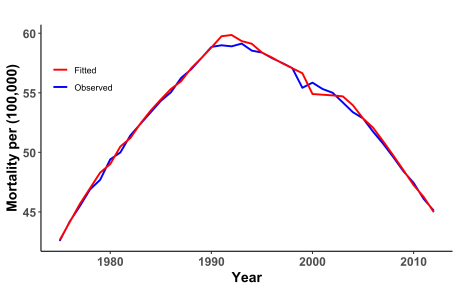}
\caption{Observed and fitted values using HDES-ARIMA}
\label{fig:4}
\end{figure}

\begin{figure}[htbp!]
\centering
\includegraphics[width=0.5\textwidth]{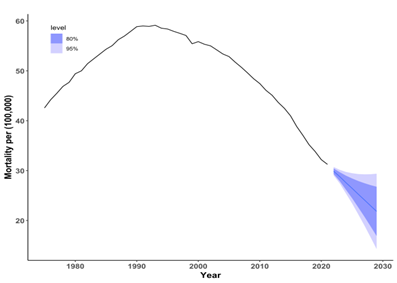}
\caption{Forecast (2022-2030) using the HDES-ARIMA}
\label{fig:5}
\end{figure}

\noindent Additionally, the HDES model parameters $\alpha$ and $\beta$ estimates were 0.5221 and 0.5219 respectively via the grid search using the training data. The plot of the residual analysis in Figure \ref{fig:3} shows that the residuals are white noise and are normally distributed. Statistical tests on the residuals indicate no deviations: Shapiro-Wilk (normality, p-value = 0.431), and Box-Ljung (white noise, p-value = 0.523). The SES model's smoothing parameter $\alpha$ was also estimated as 0.999 via grid search. However, residual analysis for the SES model showed significant deviations from the assumptions of white noise but not normality: Box-Ljung (white noise, p-value $<$ 0.001) and Shapiro-Wilk (normality, p-value = 0.09178). Cross-validation on the test data revealed that both ARIMA (0,2,2) and HDES models had the lowest RMSE of 2.56. An average model (HDES-ARIMA) combining ARIMA and HDES was also constructed, yielding the same RMSE as the individual models. Figure \ref{fig:4} shows a plot of the observed values and the fitted values using the average model while Figure \ref{fig:5} shows the forecast trend for the next 8 years (2022-2030). The summary of cross-validation results is provided in Table \ref{Tab:2}.

\begin{table}[h!]
\centering
\caption{Parameter Estimates and Cross-Validation Metrics for SES, HDES, ARIMA, and the Combined HDES-ARIMA Model.}
\label{Tab:2}
\begin{tabular}{lcccr}
\toprule
Models & Parameter Estimates & RMSE & MAE & MAPE \\
\midrule
ARIMA & (0,2,2) & 2.56 & 2.16 & 6.29 \\
SES & $\alpha$ = 0.999 & 8.90 & 7.82 & 22.6 \\
HDES & $\alpha$ = 0.52, $\beta$ = 0.52 & 2.56 & 2.16 & 6.28 \\
HDES-ARIMA & $\alpha$ = 0.52, $\beta$ = 0.52, p = 0, d, q = 2 & 2.56 & 2.16 & 6.29 \\
\bottomrule
\end{tabular}
\end{table}

\section{Discussion}

\noindent This study applied three time series models ARIMA, SES, and HDES to lung cancer mortality rates in the USA utilizing SEER data. This study tries to compare and build on the work of Altuhaifa (2023) by expanding the number of years from 2018 to 2021. The Ljung Box test on the residuals of ARIMA (0,2,2) which was selected as the best among several fitted ARIMA models based on its low AIC value shows that the model was not affected by autocorrelation and as such the chosen values of the parameters p, d, and q are appropriate for the data. This result is consistent with that of Altuhaifa (2023) \cite{altuhaifa2023}.

\noindent The fitted SES model with smoothing parameter $\alpha$ whose value is estimated as 0.999 did not satisfy the residual assumptions of normality and white noise. This, however, is expected as the SES model does not capture the trend in the data. The result of this model is also consistent with that of Altuhaifa (2023) \cite{altuhaifa2023}.

\noindent Holt's Double Exponential Smoothing (HDES) model which is an extension of the Simple Exponential Smoothing (SES) model by accounting for the trend component in the data was also fitted and its parameters $\alpha$ and $\beta$ values were estimated as 0.5521 and 0.5219 respectively. The Ljung Box test on the residuals shows that they are white noise, and the Shapiro-Wilk test confirmed that residuals were normally distributed. The parameter estimates obtained in this study differ from that of Altuhaifa (2023) \cite{altuhaifa2023}. This can be attributed to the differences in the length of the data and the differences in the adjustment of the mortality rates.

\noindent Model evaluations for the three models using RMSE, MAE, and MAPE as the evaluation metrics indicated that the SES model generated the highest RMSE of 8.90. This is consistent with the findings of Altuhaifa (2023) \cite{altuhaifa2023}. The ARIMA model and the HDES model yielded the same RMSE value of 2.56. The small values of RMSE of this study as compared to the large RMSE values Altuhaifa (2023) can be attributed to the differences in the training and test data \cite{altuhaifa2023}.

\noindent While ARIMA models generally perform well with large historical data, the HDES model is faster to implement and better for short-term forecasting on large datasets when seasonality is not a concern. An average model consisting of these two models was proposed and fitted for forecasting accuracy.

\noindent The average model denoted as HDES-ARIMA, also upon evaluation using the test data yielded the same RMSE of 2.56 as the individual models used in constructing it. While the average model has the same RMSE value as the individual models, the forecast produced by this model may be more accurate compared to that of the individual models in the long run.

\noindent Based on the average model's forecasts, lung cancer mortality rates are decreasing, which may be explained by fewer smokers, early identification, and better therapies in recent years. A recent, thorough Global Burden of Disease analysis study supports the finding that mortality rates are dropping in the United States \cite{healthdataBurdenDiseases}.

\section{\Large Conclusion}

\noindent According to the Centers for Disease Control and Prevention (CDC), lung cancer is the third leading form of cancer in the USA, and every year the government spends millions of dollars on cancer research \cite{cdcLungCancer}. The National Cancer Institute (NCI) allocated \$477.4 million to lung cancer research in 2023 \cite{cancer2024Budget}. In total, Congress allocated \$7.104 billion to the NCI in 2023, plus an additional \$216 million from the 21st Century Cures Act \cite{cancer2024Budget}. Accurately forecasting lung cancer mortality is crucial for allocating healthcare resources effectively and implementing preventive strategies.

\noindent This study found that both the HDES and ARIMA models, along with an averaged combination of the two, yielded reliable downward forecasts of lung cancer death rates, with comparable performance in terms of RMSE and adherence to model assumptions. The downward trend in lung cancer mortality rates indicates a promising future in the fight against this fatal disease. Our thorough investigation shows that the combined impacts of lower smoking prevalence, improved screening procedures, and more effective treatments are driving a long-term drop in lung cancer fatalities. The models' forecasts demonstrate how public health measures aimed at reducing smoking have had an apparent influence on lowering overall mortality rates. The anticipated drop reflects the successful application of early detection techniques, particularly current imaging technologies and stringent screening protocols. These findings illustrate the positive effects of diverse approaches to lung cancer treatment that include prevention, early intervention, and medicinal breakthroughs.

\noindent These results imply that continuing adherence to tobacco control policies, together with emerging treatment options, will most likely maintain the positive drop in lung cancer fatalities. The forecasting results demonstrate the effectiveness of current public health policies while also encouraging the pursuit of new advances in cancer detection and treatment. These projections provide hope that the combined effect of behavioral change, screening innovations, and advances in medicine will continue to lower the burden of lung cancer. The models' findings demonstrate that a long-term multifaceted approach to lung cancer control can result in significant reductions in mortality rates.

\noindent A future study can expand the dataset to include monthly rather than only annual data, which may improve forecasting accuracy and provide holistic information of short-term trends of lung cancer mortality. Additionally, future research could benefit from disaggregating forecasts by gender, ethnicity, and state to better reveal intrinsic dynamics of lung cancer mortality trends among the diverse groups of people with lung cancer in the US. This approach may support more tailored public health policies and interventions.

\vspace{1cm}

\section*{Acknowledgment}
\vspace{-0.4cm}
The lead author extends sincere gratitude to all co-authors for their collaborative efforts, intellectual contributions, and support throughout the research process.
\vspace{-0.5cm}
\section*{Conflict of Interest}
\vspace{-0.5cm}
The authors declare no conflict of interest.
\vspace{-0.5cm}

\section*{Funding}
\vspace{-0.5cm}
This research received no specific grant from any funding agency in the public, commercial, or not-for-profit sectors.
\vspace{-0.5cm}

\newpage

\printbibliography

\end{document}